\documentclass[twoside]{article}
\usepackage{fleqn,espcrc2}
\usepackage[dvips]{graphicx}
\usepackage{amsmath}

\newcommand{\kstar}{\ensuremath{K^{*}(892)^0}}
\newcommand{\khigh}{\ensuremath{K^{*}(1680)^0}}
\newcommand{\klow}{\ensuremath{K^{*}_{0}(1430)^0}}
\newcommand{\ktensor}{\ensuremath{K^{*}_{2}(1430)^0}}

\newcommand{\kstarb}{\ensuremath{\overline{K}{}^{*}(892)^0}}
\newcommand{\khighb}{\ensuremath{\overline{K}{}^{*}(1680)^0}}
\newcommand{\klowb}{\ensuremath{\overline{K}{}^{*}_{0}(1430)^0}}

\newcommand{\Dkstarmunu}{\ensuremath{D^+ \!\rightarrow\! \kstarb \mu^+ \nu}}
\newcommand{\Dkhighmunu}{\ensuremath{D^+ \!\rightarrow\! \khighb \mu^+ \nu}}
\newcommand{\Dklowmunu}{\ensuremath{D^+ \!\rightarrow\! \klowb \mu^+ \nu}}

\newcommand{\Dkpipi}{\ensuremath{D^+ \!\rightarrow\! K^- \pi^+ \pi^+}}
\newcommand{\Dkpimunu}{\ensuremath{D^+ \!\rightarrow\! K^- \pi^+ \mu^+ \nu }}
\newcommand{\mkpi}{\ensuremath{m_{\textrm{K$\pi$}}}}

\def\etal{et al.}

\newcommand{\thv}{\ensuremath{\theta_\textrm{v}}}
\newcommand{\thl}{\ensuremath{\theta_\ell}}

\newcommand{\qsq}{\ensuremath{q^2}}

\newcommand{\prd}[1]{Phys.~Rev.~D \textbf{#1}}
\newcommand{\plb}[1]{Phys.~Lett.~B \textbf{#1}}
\newcommand{\prl}[1]{Phys.~Rev.~Lett. \textbf{#1}}
\newcommand{\zpc}[1]{Z.~Phys.~C \textbf{#1}}
\newcommand{\npb}[1]{Nucl.~Phys.~B \textbf{#1}}

\title{Analysis of the $K\pi$ hadronic state interaction using \Dkpimunu{} semileptonic decays from the FOCUS experiment}
\author{A. Massafferri \address{Centro Brasileiro de Pesquisas F\'\i sicas, Brasil} \thanks{Now at University of Rome ``Tor Vergata'' and INFN, Rome, Italy} on behalf of the FOCUS Collaboration}

\begin{document}

\begin{abstract}

We present a four-body semileptonic charm decay \Dkpimunu{} analysis in the range of 0.65 GeV/$c^2$ $< \mkpi <$ 1.5 GeV/$c^2$. We observe a low mass scalar contribution of $5.30 \pm 0.74^{\,+\,0.99}_{\,-\,0.96}\%$ with respect to the total \Dkpimunu{} decay, compatible with the phase shift found by the LASS elastic scattering experiment. For the \kstar{} resonance, we obtain a mass of $895.41 \pm 0.32^{\,+\,0.35}_{\,-\,0.43}$ MeV/$c^2$, a width of $47.79 \pm 0.86^{\,+\,1.32}_{\,-\,1.06}$ MeV/$c^2$, and a Blatt-Weisskopf damping factor parameter of $3.96 \pm 0.54^{\,+\,1.31}_{\,-\,0.90}$ GeV$^{-1}$. We also report 90\% CL upper limits of 4\% and 0.64\% for the branching ratios $\frac {\Gamma (D^+ \to \khighb \mu^+ \nu)} {\Gamma (D^+ \to K^- \pi^+ \mu^+ \nu)}$ and $\frac{\Gamma (D^+ \to \klowb \mu^+ \nu)} {\Gamma (D^+ \to K^- \pi^+ \mu^+ \nu)}$, respectively.

\end{abstract}

\maketitle

\section{Introduction}

Weak semileptonic decays of charm mesons continue to attract interest due to the relative simplicity of their theoretical description: the matrix element of these decays can be factorized as the product of the leptonic and hadronic currents. This makes the \Dkpimunu{} decay a natural place to study the $K\pi$ system in the absence of interactions with other hadrons. 

It is known that the $K\pi$ final state of \Dkpimunu{} decay is strongly dominated by the \kstar{} vector resonance \cite{kstardominance,e687}. The large and clean sample of $D^+\!\rightarrow\! K^-\pi^+\mu^+\nu$ events collected by the Fermilab FOCUS experiment provides an excellent opportunity to measure the \kstar{} mass and width, as well as the effective Blatt-Weisskopf damping factor parameter discussed in Ref. \cite{blatt}. 

\begin{figure}
\begin{center}
\vspace{-0.4cm}
\includegraphics[width=9cm,height=6cm]{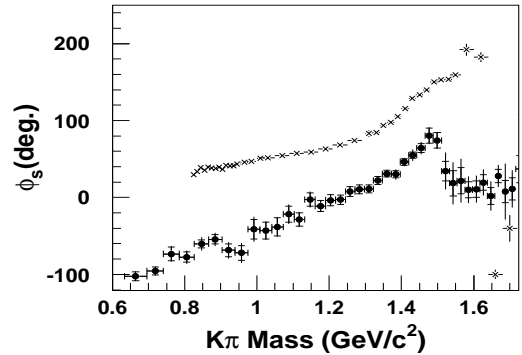}
\vspace{-1cm}
\caption{Comparisons between the results obtained for the s-wave $K\pi$ scattering phase shift measured by LASS and the one measured by the E791 experiment (solid circles) in \Dkpipi{} charm hadronic decay (Figure provided by B. Meadows).
\label{Lass_Kpipi}}
\end{center}
\end{figure}

In addition we also search for structures other than the \kstar{} resonance in the mass range of 0.65 GeV/$c^2$ $< \mkpi{} <$ 1.5 GeV/$c^2$. The available phase space allows us to investigate the low mass scalar sector whose spectroscopy has been increasingly studied since the indication of the existence of the broad scalar resonance $\kappa$ in \Dkpipi{} decays from a Dalitz Plot analysis performed by E791~\cite{dkpipi}. Recently E791 has also pursued an independent approach for the scalar component extracting directly the scalar phase shift from their data~\cite{brian}. 
These two results, obtained using \Dkpipi{} hadronic decay is substantially different from the one obtained in a partial wave analysis performed by the LASS scattering experiment~\cite{lass}. LASS observed that the s-wave amplitude in the $K^-\pi^+$ elastic region can be represented as the sum of a \klow{} resonance and a smooth shape, consistent with the non-resonant hypothesis. Figure \ref{Lass_Kpipi} shows a comparison of the phase shift of the scalar component found by LASS to the one found by E791 in \Dkpipi{} decay. 
%The two E791 results, obtained from \Dkpipi{} hadronic decays are substantially different from the results obtained in a partial wave analysis performed by the LASS scattering experiment.  While the E791 analysis required a $\kappa$, LASS observed that the s-wave amplitude in the $K^-\pi^+$ elastic region can be represented as the sum of a \klow{} resonance and a smooth shape, consistent with the non-resonant hypothesis.  The scalar phase shift observed by LASS and E791 are compared in Fig. \ref{Lass_Kpipi}.

Due to Watson's Theorem~\cite{bigi,watson} the observed $K\pi$ phase shifts in \Dkpimunu{} decays should be the same as those measured in $K\pi$ elastic scattering. This establishes a connection between the production and the scattering mechanisms, adding crucial information about $K\pi$ hadronic interaction and the role of the bachelor $\pi$ final state interaction in hadronic decays. 

\section{The Experiment and the Data Sample} 

The data were collected in the Wideband photoproduction experiment FOCUS during the Fermilab 1996--1997 fixed-target run. In FOCUS, a forward multi-particle spectrometer is used to measure the interactions of high energy photons on a segmented BeO target. The FOCUS detector is a large aperture, fixed-target spectrometer with excellent vertexing and particle identification. The FOCUS beamline~\cite{beam}, the detector~\cite{topher1,detector1,detector2,detector3} and the specific cut selection~\cite{mypaper} have been described elsewhere. A total of 18245 \Dkpimunu{} candidates remain after the selection criteria. We estimate the charm and non-charm background contributions to be, respectively, 17.8\% and 3.2\% of the total number of events over our signal region.

\section{The \Dkpimunu{} signal description} 

Four-body decays of spinless particles are described by five kinematic variables. The variables chosen in this analysis are the $K^-\pi^+$ invariant mass (\mkpi{}), the square of the $\mu^+\nu$ mass (\qsq{}), and three decay angles: the angle between the $\pi^+$ and the $D^+$ direction in the $K^-\pi^+$ rest frame (\thv{}), which defines one decay plane, the angle between the $\nu$ and the $D^+$ direction in the $\mu^+\nu$ rest frame (\thl{}), which defines the second decay plane, and the acoplanarity angle ($\chi$) between these two decay planes. 
The differential decay rate can be represented by a coherent sum of resonant and non-resonant contributions to the angular momentum eigenstates of the $K^-\pi^+$ system, 
\begin{equation}
\frac{d\Gamma}{d m_{K\pi}} = \int {\left| \sum_{J} \sum_{R} a_{J,(R)} ~\mathcal{M}_{J} ~\mathcal{A}_{J,(R)} \right|^2 }~\phi ~d\Omega
\label{eq1}
\end{equation}

where $d\Omega \equiv d\qsq~d\!\cos{\thv}~d\!\cos{\thl}~d\chi$, $\mathcal{M}_{J}$ is the weak matrix element for a transition with angular momentum $J$, $\mathcal{A}_{J,(R)}$ represents the form of the hadronic final state amplitude contribution of resonance $R$ (or non-resonant) with strength $a_{J,(R)}$, and $\phi$ is the phase space density. 

The possible resonant states that couple to $K^-\pi^+$ are the scalars $\kappa$ and \klow{}, the vectors \kstar{} and \khigh{}. \footnote{Due to the orthogonality of states with different angular momentum, only amplitudes with the same spin will produce significant interference contributions to the \mkpi{} mass spectrum. The inclusion of a small \ktensor{} resonance contribution is unlikely to be observed, since it is orthogonal to the (dominant) \kstar{} and low mass s-wave amplitudes.} The non-resonant contribution is assumed to be scalar.

The parametrization of resonant states with angular momentum $J$ is given by the product of a Breit-Wigner and the normalized $R \to K^-\pi^+$ coupling, $\mathcal{F}_{J}$

\begin{equation}
\mathcal{A}_{J,R} = \frac{m_0\,\Gamma_0}{m_{K\pi}^2 - m_0^2 + i\,m_0\,\Gamma(m_{K\pi})}\,\mathcal{F}_{J} 
(m_{K\pi})
\label{eq2}
\end{equation}

where $\Gamma(m_{K\pi}) = \Gamma_0\,\mathcal{F}_{J}^2\,\frac{p^*}{p^*_0}\,\frac{m_0}{m_{K\pi}}$, $p^{*}$ is the magnitude of the kaon momentum in the resonance rest frame, $p^{*}_0 = p^{*}(m_{0})$, $\mathcal{F}_0 = 1$, and $\mathcal{F}_1 = \frac{p^*}{p^*_0}\,\frac{B(p^{*})}{B(p^{*}_0)}$. $B$ is the Blatt-Weisskopf damping factor given by $B = 1 / \sqrt{1 + r_0^2 ~p^{*2}}$~\cite{blatt}. The damping factor adds an additional fit parameter, $r_0$, in our fits to the \kstar{} line shape. The line shape of the $\kappa$ resonance is expected to deviate significantly from a pure Breit-Wigner, due to its large width and the close vicinity of the $K\pi$ threshold. In this analysis we use the $\kappa$ line-shape adopted by E791~\cite{dkpipi} while for the non-resonant component we use an empirical phase shift, parametrized by the Effective Range model, obtained by LASS (Fig. \ref{Lass_Kpipi}) \footnote{We have to remove the two-body phase space factor, given by $\frac{p^*}{m_{K\pi}}$, from LASS non-resonant amplitude, which is already included in Eq.~\ref{eq1}.}.

The weak matrix element for the vector process, $\mathcal{M}_1$, and for the scalar process, $\mathcal{M}_0$, are written as a function of helicity amplitudes, $H_{i}$, derived in~\cite{schuler}. 

\section{Angular Distribution Results} 

\begin{figure}[tbp]
\begin{center}
\includegraphics[width=7cm,height=5cm]{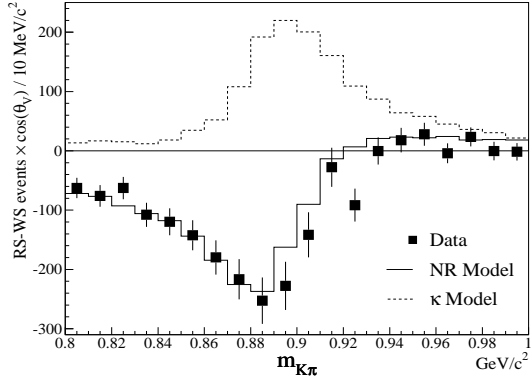}
\vspace{-1cm}
\caption{
The background subtracted distribution of \mkpi{} weighted by $\cos{\thv{}}$. The data (squares) show good agreement with the LASS non-resonant parameterization (solid histogram) but not with a $\kappa$ model (dashed histogram)}
\label{fig_asymmetry}
\vspace{-1cm}
\end{center}
\end{figure}

Next we discuss the angular distribution described by Eq.~\ref{eq1}. The $K\pi$ spectrum described by this equation includes the dominant contribution from the \kstar{} resonance, possible high mass contributions from the \klow{} and \khigh{} resonances, and low mass scalar components comprised of a non-resonant and a possible $\kappa$ contributions, both populating the region where significant discrepancies were found between the data and the predicted \Dkstarmunu{} angular decay distributions. It has been shown in ~\cite{topher1,topher2} that a nearly constant amplitude and phase contribution to the helicity zero amplitude of the virtual $W^+$ was required to adequately fit the observed decay angular distributions. The \mkpi{} distribution weighted by $\cos{\thv{}}$ provides information on the phase of the additional structure relative to that of the \kstar{}. It can be used to discriminate different combinations of low mass states, given the large difference between their phase shifts. Figure~\ref{fig_asymmetry} compares the distribution obtained in the data with the predictions from the non-resonant and $\kappa$ models in the absence of additional phase shifts as required by Watson's Theorem. Since a simulation using the LASS parametrization of the non-resonant contribution is sufficient to reproduce the data, we exclude a possible $\kappa$ contribution from further consideration.

\section{$K\pi$ Mass Spectrum Analysis}

Having excluded the $\kappa$, the most general differential decay rate for \Dkpimunu{} in \mkpi{} is given by Eq.~\ref{eq5b}. 

\begin{equation}
\frac{d\Gamma}{d m_{K\pi}} = |\mathcal{V}|^2 F_{11} + |\mathcal{S}|^2 F_{00} + 2\,\Re (\mathcal{V^*~S}) F_{01} 
\label{eq5b}
\end{equation}

The $F_{JJ^{'}} \equiv \int \epsilon \mathcal{M}^*_{J} \mathcal{M}_{J^{'}}\,\phi\,d\Omega$, are real functions\footnote{All imaginary pieces of $\mathcal{M}^*_{J} \mathcal{M}_{J^{'}}$  will appear as sinusoidal functions of $\chi$. Hence any imaginary terms vanish when averaged over $\chi$ given our nearly uniform acceptance in this variable.} that depend only on \mkpi{}. These functions are computed from the \mkpi{} spectrum obtained from a complete simulation of \Dkpimunu{} events, generated according to phase space and weighted by $\mathcal{M}^*_{J} \mathcal{M}_{J^{'}}$ and thus represent the intensity modified by acceptance and efficiency, $\epsilon$. The vector, $\mathcal{V}$, and scalar, $\mathcal{S}$, amplitudes are given by the sum of the relevant states, each one represented by the product of a real magnitude, $a$, and $\mathcal{A}$. The $|\mathcal{V}|^2$, $|\mathcal{S}|^2$, and  $\Re (\mathcal{V}^* \mathcal{S})$ functions depend on \mkpi{} as well as on all fit parameters. The cross-term, $2\,\Re (\mathcal{V^*~S}) F_{01}$, represents the interference between the vector and scalar contributions. 

The fit parameters are the magnitudes of each amplitude, $a_i$, the mass and width of the \kstar{}, and the parameter $r_0$ of the Blatt-Weisskopf damping factor. The parameters of all other resonances are fixed to the PDG values~\cite{pdg}. The fit parameters are obtained from an unbinned maximum-likelihood method, by minimizing the quantity  $- 2 \ln{\sum_{\textrm{events}}\left[ (1 - f_{B}) \mathcal{L}_{S} + f_{B} \mathcal{L}_{B} \right]}$, where $\mathcal{L}_{S}$ is the probability density for the signal, $\mathcal{L}_{B}$, for the background, and $f_B$ is the background fraction, fixed in the estimated values. The \kstar{} amplitude is taken as the reference amplitude. Decay fractions are obtained integrating each individual amplitude over the phase space and dividing by the integral over the phase space of the overall amplitude.

To account for momentum resolution effects on the \kstar{} parameters, we refit the data fixing all parameters except the \kstar{} width and use a probability density function convoluted with a Gaussian distribution with $\sigma = 5.88~\textrm{MeV}/c^2$, value obtained from Monte Carlo simulation. The smearing due to momentum resolution increases the \kstar{} width by approximately $2~\textrm{MeV}/c^2$. 

\section{$K\pi$ Mass Spectrum Results}

Using the procedure described above, we fit the data assuming only a \Dkstarmunu{} process. The confidence level of this fit is 0.21\%, indicating the need for additional contributions in the decay.

\begin{figure}
\begin{center}
\includegraphics[width=7cm,height=6cm]{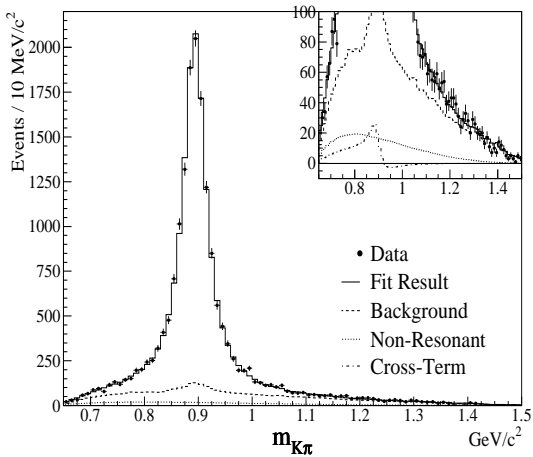}
\vspace{-1cm}
\caption{Fit to the \mkpi{} data using the NR model. The error bars, the solid lines, the dashed lines, and the dotted lines correspond to the data, the model, the background contribution, and the scalar contribution, respectively. The upper right plot shows the same information and the cross-term (dot-dash line) with a limited y-axis to allow more detail to be seen. 
\label{fig_results}}
\vspace{-1cm}
\end{center}
%\end{figure*}
\end{figure}

The inclusion of a non-resonant scalar component, referred to as the \emph{NR model}, significantly improves the confidence level of the fit to 66\%.  We find $m_{K^{*}(892)^0} = 895.41 \pm 0.32$~MeV/c$^2$, $\Gamma_{K^{*}(892)^0} = 47.79 \pm 0.86$~MeV/c$^2$, $r_0 = 3.96 \pm 0.54$~GeV$^{-1}$ and $a_{\textrm{NR}} = 0.327 \pm 0.024$, which correspond to a scalar fraction of $5.30 \pm 0.74$ \%. Figure~\ref{fig_results} illustrates the contribution of both the \Dkstarmunu{} and non-resonant s-wave process to the observed \mkpi{} spectrum. 

%\begin{figure*}

We also consider \Dkhighmunu{} and \Dklowmunu{} contributions to our model. The observed \khigh{} and \klow{} amplitudes are consistent with zero and we find $\frac {\Gamma (D^+ \! \to \! \khighb \mu^+ \nu)} {\Gamma (D^+ \! \to \! K^- \pi^+ \mu^+ \nu)} < 4.0$\% and $\frac{\Gamma (D^+ \! \to \! \klowb \mu^+ \nu)} {\Gamma (D^+ \! \to \! K^- \pi^+ \mu^+ \nu)}<0.64$\% at 90\% CL. The upper limits are calculated using the method described in~\cite{lions} and assume $\textrm{BR}(\khighb \!\to\! K^-\pi^+)=0.258$ and $\textrm{BR}(\klowb \!\to\! K^-\pi^+)=0.62$~\cite{pdg}. To study the statistical significance of these new amplitudes, we use a hypothesis test based on the maximum-likelihood ratio method~\cite{testhypothesis}. As a result, we obtain a confidence level of 80\% in favor of the simple NR model. 

\section{Summary and Discussion}
       
In conclusion we find that our angular distribution is consistent with the effective-range scalar non-resonant phase shift obtained by LASS~\cite{lass} as expected by Watson's Theorem given the absence of other final state interactions (FSI). It points out that the FSI of the bachelor pion of the $D \to (K\pi)\pi$ decay can play an important role towards the observed difference with respect to $K\pi$ scalar scattering phase shift. 

We have measured the \kstar{} parameters using a large sample of \Dkstarmunu{} signal events over a wide mass range. The absence of high mass resonances as well as the small background contribution provides a unique environment to study the \kstar{} mass and width. Our measurements of the mass and width are more than $1\,\sigma$ below the present world average value. We obtain a Blatt-Weisskopf parameter consistent with the value obtained by LASS~\cite{lass}. We also limit possible additional $K\pi$ resonances present in \Dkpimunu{} semileptonic decays. 

\section{Acknowledgments}

We wish to acknowledge the assistance of the staffs of Fermi National Accelerator Laboratory, the University of Rome ``Tor Vergata'', the INFN of Italy, the Universidade Federal do Rio de Janeiro, the Funda\c c\~ao de Amparo \`a Pesquisa do Estado do Rio de Janeiro and the physics departments of the collaborating institutions. This research was supported in part by the U.~S. National Science Foundation, the U.~S. Department of Energy, the Italian Instituto Nazionale di Fisica Nucleare and Ministero dell'Universit\`a e della Ricerca Scientifica e Tecnol\'ogica, the Brazilian Conselho Nacional de Desenvolvimento Cient\'{\i}fico e Tecnol\'ogico, CONACyT-M\'exico, the Korean Ministry of Education, and the Korea Research Foundation.

\end{document}